\documentclass[preprint,review,12pt]{elsarticle}

\usepackage[margin=2.65cm]{geometry}% for document class: preprint

\usepackage[noend]{algpseudocode}
\usepackage[english]{babel}
\usepackage{verbatim}
\usepackage[table,xcdraw]{xcolor}
\usepackage{graphicx} 
\usepackage{amssymb}
\usepackage{amsmath}
\usepackage{amsfonts}
\usepackage{mathrsfs}
\usepackage{mathtools}
\usepackage{float}
\usepackage{multirow}
\usepackage{tabularx}
\usepackage{array}
\usepackage{color}
\usepackage{leftidx}
\biboptions{numbers,sort&compress}
\usepackage{xcolor}
\usepackage[]{algorithm2e}
\usepackage{textgreek}
\usepackage[colorlinks,citecolor=red,urlcolor=blue,bookmarks=false,hypertexnames=true]{hyperref} 
\usepackage{multicol}%
\usepackage{comment}

\usepackage{booktabs}
\usepackage{caption}
\usepackage{subcaption}

\usepackage{todonotes}
\definecolor{MC}{rgb}{0.0, 0.5, 0.0}

\newcommand*{\colorboxed}{}
\def\colorboxed#1#{%
  \colorboxedAux{#1}%
}
\newcommand*{\colorboxedAux}[3]{%
  \begingroup
    \colorlet{cb@saved}{.}%
    \color#1{#2}%
    \boxed{%
      \color{cb@saved}%
      #3%
    }%
  \endgroup
}

\usepackage{dashbox}

\usepackage{tikz}
\usepackage{soul}

\usepackage{rotating}

\usepackage{lineno}
\newcommand{\RN}[1]{%
  \textup{\uppercase\expandafter{\romannumeral#1}}%
}

\definecolor{azure(colorwheel)}{rgb}{0.0, 0.5, 1.0}
\definecolor{mygreen}{rgb}{0.0, 0.5, 0.0}
\definecolor{darkraspberry}{rgb}{0.53, 0.15, 0.34}
\definecolor{bleudefrance}{rgb}{0.19, 0.55, 0.91}

\newcolumntype{C}[1]{>{\centering\arraybackslash}m{#1}}

\newcommand{\eg}{\textit{e}.\textit{g}., }
\newcommand{\ie}{\textit{i}.\textit{e}., }

\begin{document}

\begin{frontmatter}

% \dochead{}

\title{Augmentation of scarce data - a new approach for deep-learning modeling of composites}

%Navigating data scarcity: a novel data augmentation approach for deep-learning in composite modeling
%Expanding limited data: data augmentation for deep-learning modeling of composites
%An original data augmentation approach to produce unoriginal data for deep-learning modelling of composites
%A novel data augmentation approach for deep-learning modeling of composites
%An efficient data augmentation approach for deep-learning modeling of composites with scarce data

\author[mymainaddress]{Hon Lam Cheung}
\author[mysecondaddress]{Petter Uvdal}
\author[mysecondaddress]{Mohsen Mirkhalaf\corref{mycorrespondingauthor}}

\cortext[mycorrespondingauthor]{Email: mohsen.mirkhalaf@physics.gu.se}

\address[mymainaddress]{Department of Mechanics and Maritime Sciences, Department of Mechanics and Maritime Sciences, Hörsalsvägen 7A, 41296 Gothenburg, Sweden}
\address[mysecondaddress]{Department of Physics, University of Gothenburg, Origovägen 6B, 41296 Gothenburg, Sweden}

\begin{abstract}
High-fidelity full-field micro-mechanical modeling of the non-linear path-dependent materials demands a substantial computational effort. Recent trends in the field incorporates data-driven Artificial Neural Networks (ANNs) as surrogate models. However, ANNs are inherently data-hungry, functioning as a bottleneck for the development of high-fidelity data-driven models. This study introduces a novel approach for data augmentation, expanding an original dataset without additional computational simulations. A Recurrent Neural Network (RNN) was trained and validated on high-fidelity micro-mechanical simulations of elasto-plastic short fiber reinforced composites. The obtained results showed a considerable improvement of the network predictions trained on expanded datasets using the proposed data augmentation approach. The proposed method for augmentation of scarce data may be used not only for other kind of composites, but also for other materials and at different length scales, and hence, opening avenues for innovative data-driven models in materials science and computational mechanics.
\end{abstract}


\end{frontmatter}
%#########################################################################################
\section{Introduction}
\label{Introduction}
Classical high-fidelity numerical simulations of the elasto-plastic response of Short Fiber Reinforced Composites (SFRCs) require Finite Element (FE)- or Fast Fourier Transform (FFT)-analysis, which demands a substantial computational effort \cite{Schneider2016}. Since, numerical simulations are a crucial part to an iterative design process, there is a need for faster, yet accurate, computational modeling of the non-linear response of SFRCs \cite{Spahn2014}. However, data-scarcity of high-fidelity simulations remains a limiting factor for accurate data-driven models. Therefore, in this study, we have developed a data augmentation approach for deep-learning of composites, with the application of a surrogate model for the elasto-plastic response of SFRCs using high-fidelity data. 

 SFRCs have promising lightweight applications across various industries, including aerospace, automotive, marine, and civil engineering applications \cite{Hagnell2019, Jain2016, Mortazavian2015}. These materials possess a high strength and  ratios (compared to unfilled matrices) and can be manufactured quickly and at low-cost using injection molding, enabling the material to form complex 3D shapes \cite{Mortazavian2015}. Specific applications, \eg inelastic energy absorption during high impacts \cite{Tikarrouchine2018}, require an understanding of the materials non-linear elasto-plastic response. The macro-mechanical response of an SFRC is highly dependent on a wide variety of micro-structural parameters, including morphological and constitutive properties \cite{Selmi2011,Tian2015,Mirkhalaf2020}.

To quantitatively predict the behavior of SFRCs, micro-mechanical models, including low-fidelity mean-field models or high-fidelity full-field models, are used \cite{Mirkhalaf2020}. Full-field models analyzes a Representative Volume Element (RVE) to obtain a homogenized material response by volume averaging, using FE \cite{Tian2015, Tikarrouchine2018, Qi2015} or FFT \cite{Spahn2014, Schneider2016}. These models have a strong predictive capability. However, simulations are computationally expensive and RVE generation is challenging when it comes to high fiber aspect ratios and high fiber volume fractions \cite{Hoang2016, Harper2012, Bargmann2018, Pan2008, Mirkhalaf2019-ICCM}. As a result, despite recent advances in the field, modeling non-linear path-dependent SFRCs still requires a considerable computational effort.

Recently, there have been developments of data-driven modeling of composites using ANNs (see \eg \cite{Mozaffar2019, Mentges2021, LIU2021, Friemann2023, Ghane2023, Bonatti2022, Liu2023, Cheung2024}). ANNs offer several advantages over conventional models as they enable fast and efficient calculations of complex correlations \cite{Lecun2015, Agrawal2019, Bonatti2022}. To model plasticity, Mozaffar et al. \cite{Mozaffar2019} implemented a Recurrent Neural Network (RNN) model, and accurately predicted the non-linear path-dependent behavior in a 2D model. One shortcoming of such an RNN is the limited long-term memory. Instead, Bonatti and Mohr developed a self-consistent RNN, the response of which did not depend on the path-sampling size, \ie it was independent of the number of strain-increment \cite{Bonatti2022}. Moreover, there have been developments in mechanistically informed neural networks to ensure coherence with the laws of physics \cite{LIU2021, Liu2023}. Liu et al. \cite{Liu2023} developed a physics-based neural network model in which the surrogate model is composed of two ANNs: one which calculates the yield and the elastic response, and a second ANN for the non-linear plasticity response. The recent advances of ANN architectures have solved certain challenges, however, the RNNs rely on a large training dataset, which is considered a bottleneck for training accurate RNN-based constitutive models \cite{LIU2021}.

%Since a large amount of data remains crucial for future developments in the field, various approaches have been proposed to address this issue \cite{Cheung2024, Heidenreich2023}. Computational simulations have been incorporated to cover 3D-models, even in anisotropic materials. For example, ANNs has been developed by Friemann et al. \cite{Friemann2023} and Cheung and Mirkhalaf \cite{Cheung2024} to model the path-dependent non-linear elasto-plastic response in SFRCs. Yet, high-fidelity full-field simulated data was limited due to the computational time required, in which transfer learning from mean-field to full-field has been implemented \cite{Cheung2024}. Thus it is clear that there is an increasing need of high-fidelity data, due to the data-hungry nature of ANNs.

Since a large amount of data remains crucial for future developments in the field, transfer learning approaches have been proposed to address this issue. This approach could be employed to adopt a network trained on one material and to be used for new materials with smaller additional datasets \cite{Heidenreich2023}. Recently, Ghane et al. \cite{Ghane2023-RNN} used transfer learning to address initialization challenges and sparse data issues for usage of RNNs for cyclic behavior of elasto-plastic woven composites. 
This approach can also be used to develop accurate data-driven models by initially training an ANN using a large amount of low-fidelity data, and then apply transfer learning using a small high-fidelity dataset \cite{JUNG2022,Cheung2024}. However, this approach requires multiple datasets for initial pre-training and subsequent fine-tuning of the networks. Therefore, it may not be applicable when the sources of data are limited.

In this study, we are proposing a data augmentation approach to expand an original and small high-fidelity dataset, without a need for additional simulations. This approach is particularly useful when multiple sources of data are not available, and expanding the available data is expensive, time-wise and/or cost-wise. 
The proposed approach is to consider the simulations (and/or experimental results) input and output in \textit{multiple configurations} \ie to rotate the available data from the original coordinate system to multiple coordinate systems. The proposed approach is applied to micro-mechanical FE/FFT simulated data of non-linear elasto-plastic response of SFRCs. Using the implemented data augmentation approach, we have successfully developed an RNN as a surrogate model for non-linear elasto-plastic response of SFRCs, using only a limited number of original full-field micro-mechanical simulations. 
%The new high-fidelity stress-strain path remains physically meaningful, yet is produced without additional simulations. Instead, it is produced by rotating the original high-fidelity data, \ie the second order strain path tensor, the orientation tensor of the short fibers, and the output stress path tensor. 
This is a novel approach to drastically increase a dataset size, which could be significant for future developments of ANN surrogate models not only for composites but also for other kind of materials. 

The rest of this paper is structured as follows. Section \hyperlink{section.2}{2} outlines the development of the high-fidelity original data. This includes RVE generation of SFRCs with variety of fiber orientation distributions, randomly evolving strain paths, and the homogenization of the micro-structure response using FE- or FFT-analysis. Section \hyperlink{section.3}{3} describes how the original datset was expanded using the proposed data augmentation approach. Section \hyperlink{section.4}{4} provides an in-depth explanation of the RNN set-up and its architecture. Section \hyperlink{section.5}{5} presents the results of training an RNN on various extent of rotated data and discusses the significance and limitations of the proposed approach. Section \hyperlink{section.6}{6} finishes this paper with some final remarks about the conclusions of this study.

%#########################################################################################
\section{Original Data} 
\label{Original Data}
This study relies on the original data generated by  Cheung and Mirkhalaf \cite{Cheung2024}, in which RVEs of SFRCs with specific material properties of matrix and fibers were generated for a range of different fiber orientations and fiber volume fractions. 
In addition to the material properties, 6D-strain paths were randomly generated to cover the non-linear path-dependent response. The data was generated using Digimat-FE, which solves the boundary value problem for each RVE geometry. In the following subsections, there is a detailed explanations of each step necessary to produce the original data.

\subsection{Random orientation tensor generation}
\begin{figure}[h!]
% Answer: [trim={left bottom right top},clip]
% Starting font size 14
\centering
\includegraphics[scale=0.6, trim={0 0 0 0},clip]{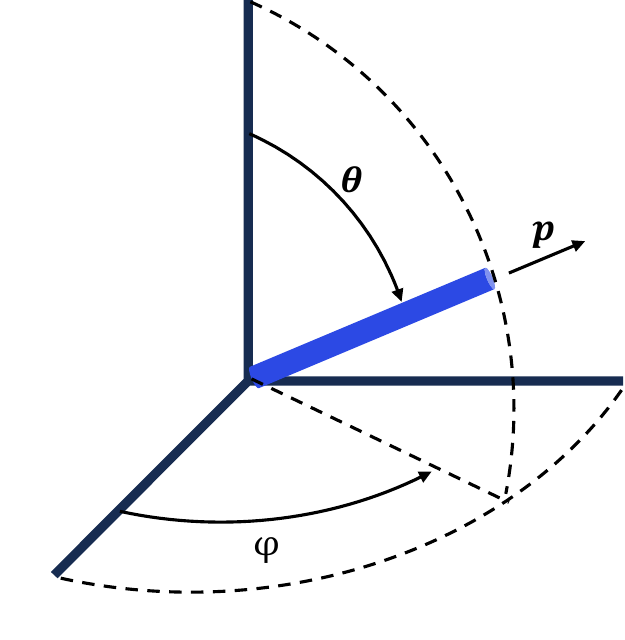}
\caption{Orientation of a short fiber ($\textbf{\textit{p}}$) in a coordinate system, described by two angles $\theta$ and $\phi$ \cite{Advani1987}.}
\label{orientation_fig}
\end{figure}

The orientation of short rigid fibers within SFRCs is influenced by the characteristics of the surrounding viscous fluid during the manufacturing process. This orientation has a significant impact on the mechanical properties of the composite. SFRCs are stronger and stiffer in the orientation of the short fibers, while more compliant in the direction of least orientation of fibers \cite{Advani1987}. Advani and Tucker \cite{Advani1987} originally described the probability distribution function of fiber orientation, denoted as  $\psi$. This distribution is related to a set of even-ordered tensors known as orientation tensors. These tensors help quantify the orientation of fibers within a composite material. 
% Fibers are assumed to be rigid cylinders with uniform length and diameter, and the concentration of fibers is assumed to be uniform throughout space. 
Fiber orientation is described using two angles, $\theta$ and  $\phi$, as shown in Figure \ref{orientation_fig}.
The fiber orientation $\textbf{\textit{p}}$ is given by
\begin{equation} \label{prob_eqn}
\boldsymbol{p}=\begin{bmatrix}
    \sin{\theta} \cos{\phi} \\
    \sin{\theta} \sin{\phi} \\
    \cos{\theta}
\end{bmatrix}.
\end{equation}
The probability $\textit{(P)}$ of a fiber direction ($\textbf{\textit{p}}$), exisiting between $\theta_{1}$ and $(\theta_{1} + d\theta$) and $\phi_{1}$ and ($\phi_1+d\phi$) is given by
\begin{equation}
    P(\theta_1 \leq \theta \leq \theta_1 + d\theta, \phi_1 \leq \phi \leq \phi_1 + d\phi) = \psi(\theta_1, \phi_1) \sin\theta_1 \, d\theta \, d\phi.
\end{equation}
The orientation distribution function $\psi(\boldsymbol{p})$ is periodic repeating with \(\pi\):
\begin{equation} \label{psi_eqn_angles}
    \psi(\theta, \phi) = \psi(\pi - \theta, \phi + \pi),
\end{equation}
\begin{equation} \label{psi_eqn_prob}
    \psi(\boldsymbol{p}) = \psi(-\boldsymbol{p}).
\end{equation}

\noindent
The integral over the unit sphere is equal to one:

\begin{equation} \label{psi_eqn_integral}
    \int_{\phi=0}^{2\pi} \int_{\theta=0}^{\pi} \psi(\theta, \phi) \sin{\theta} \,d\theta \,d\phi = \oint \psi(\boldsymbol{p}) \,d\boldsymbol{p} = 1,
\end{equation}
The components of the second order orientation tensor (\textbf{a}) is given by:
\begin{equation}
    a_{ij} = \int_{\Omega} p_i p_j \psi(\boldsymbol{p})d\boldsymbol{p}.
\end{equation}

\begin{figure}[h!]
\centering
    \includegraphics[scale=0.80, trim={0cm 0cm 0cm 0cm},clip]{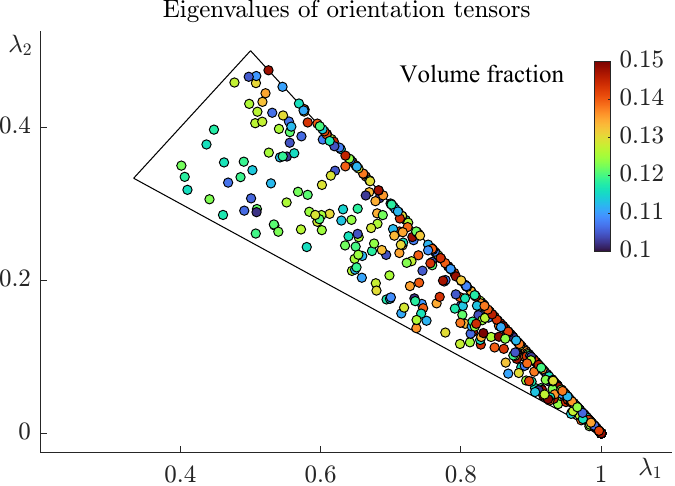}
    \caption{Eigenvalues of the randomly generated diagonal orientation tensors.}
    \label{orientation_diag}
\end{figure}

\begin{figure}[h!]
    \centering
    \includegraphics[scale=0.80, trim={15cm 5cm 1cm 4cm},clip]{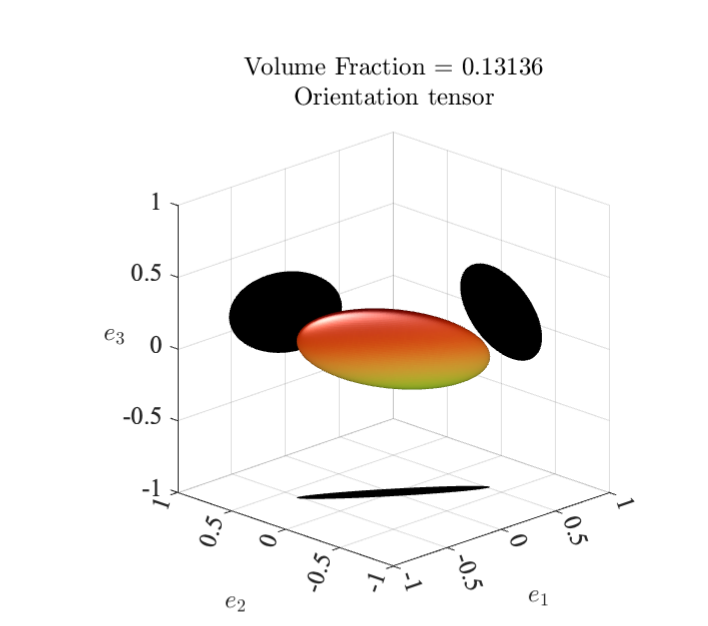}
    \caption{An example figure of randomly generated orientation tensor.}
    \label{orientation_tensor}
\end{figure}

In this study, we used the method developed by Friemann et al. \cite{Friemann2023} to generate random reference orientation tensors. For a 3D orientation tensor, eigenvalues (the components of a diagonal orientation tensor) were sampled as a set of three positive numbers summing up to one. The eigenvalues of the randomly generated orientation tensors are shown in Figure \ref{orientation_diag}. 
Subsequently, the diagonal orientation tensor were rotated to obtain the final orientation tensor, by applying a random rotation tensor generated with Arvo's algorithm \cite{Arvo1992}. An example of a randomly generated orientation tensor is shown in Figure \ref{orientation_tensor}. 
In the case of planar reference orientation tensors, one diagonal component was set to zero. Two numbers were sampled for the x and y directions, with their sum equal to one. The tensor was then randomly rotated around the z-axis, followed by a 90-degree rotation in the x, y, or z-axis. For the uni-directional orientation tensor, the first, second, or third component of the diagonal was randomly selected and set to 1, with all other components being 0.

\subsection{Random strain path generation}
Random strain paths were generated following the procedure developed by Friemann et al. \cite{Friemann2023}, resulting in a random 6D strain walk. Firstly, the number of drift directions $(\mathrm{nDrift})$ was randomly sampled from $(1,2,5,10)$. Then, each component of the 6-dimensional vector representing the drift direction was sampled individually from a normal distribution with a mean of 0 and a standard deviation of 1. Subsequently, the drift directions were normalized to have a magnitude of 1. This normalized drift direction was then repeated a set number of times (determined by the number of time steps, set as 100, divided by $\mathrm{nDrift}$). The previous steps were iterated for $\mathrm{nDrift}$ times with a new drift direction until 100 time steps was reached. For each time step, a perturbation was introduced by sampling a random noise vector from a standard distribution with a mean of 0 and a standard deviation of 1, these noise vectors were then scaled by a gamma value ranging from 0 to 1. Finally, a cumulative sum was computed by adding both the drift and noise to obtain the path. The path was then scaled so that the maximum strain equaled the specified maximum strain, which ranged from 0.01 to 0.05.

\subsection{RVE size determination and FE/FFT-simulations}
The RVE size determined by Cheung and Mirkhalaf \cite{Cheung2024}, to include sufficient microstructural detailed to accurately capture the non-linear elasto-plastic response. The approach follows the criteria proposed by Mirkhalaf et al. \cite{Mirkhalaf2016} for determining RVE size, \ie the coefficient of variation in deformation behavior should be less than a desired value; and the average responses should fall within a desirable error range. RVE sizes were selected based on a statistical analysis, which optimizes computational time with data accuracy. The approach also limits the maximum RVE size for 3D fiber orientation distributions to ensure computational efficiency. Once the RVE size was determined, FE/FFT-analysis was employed using Digimat-FE. 

It should be mentioned that a single fiber length was considered for the RVE generations. Recently, Mentges et al. \cite{Mentges2023} showed that using a Representative fiber length, accurate predictions are obtained (comparable to simulations considering a fiber length distribution). 

\subsection{Specific loading test simulations}
In addition to the random 6-dimensional loading data, specific loading tests were simulated, too. These were performed to later evaluate the effectiveness of the trained RNN on standard loading conditions. These were cyclic loading tests with the strain components going from 0 to 0.035, then to -0.035, and returning to 0. The loading cases included uniaxial normal stress (\(\sigma_{11}\)), uniaxial shear stress (\(\sigma_{12}\)), biaxial stress in two normal directions (\(\sigma_{11} + \sigma_{22}\)), biaxial stress in normal and shear (\(\sigma_{11} + \sigma_{23}\)), and a plane strain (\(\varepsilon_{11} + \varepsilon_{22}\)). Each loading test was applied to five different RVEs with random orientation tensors. Table \ref{tab:specific-test-sample-properties} provides the properties of each RVE. 

\begin{table}[h!]
\centering
\caption{Orientation tensor components, fiber volume fraction for each of the specific test samples.}
\label{tab:specific-test-sample-properties}
\scalebox{0.85}{
\begin{tabular}{cccccccc}
\hline
\textbf{Sample} & $a_{11}$ & $a_{22}$ & $a_{33}$ & $a_{12}$ & $a_{13}$ & $a_{23}$ & $v_f$  \\ \hline
\#1             & 0.0197   & 0.4926   & 0.4877   & 0.0172   & -0.0223  & 0.2963   & 0.1152 \\
\#2             & 0.5193   & 0.3062   & 0.1745   & 0.0613   & -0.1175  & -0.1690  & 0.1320 \\
\#3             & 0.6291   & 0.2851   & 0.0858   & 0.3868   & 0.0636   & 0.0443   & 0.1409 \\
\#4             & 0.7748   & 0.0946   & 0.1306   & 0.0540   & 0.1060   & -0.0576  & 0.1142 \\
\#5             & 0.4435   & 0.2642   & 0.2923   & -0.2709  & -0.0629  & 0.0234   & 0.1314 \\ \hline
\end{tabular}}
\end{table}

%#########################################################################################
\section{Data Augmentation Approach} 
\label{Data Augmentation Approach}
\begin{figure}[h!]
\centering
\includegraphics[scale=0.8]{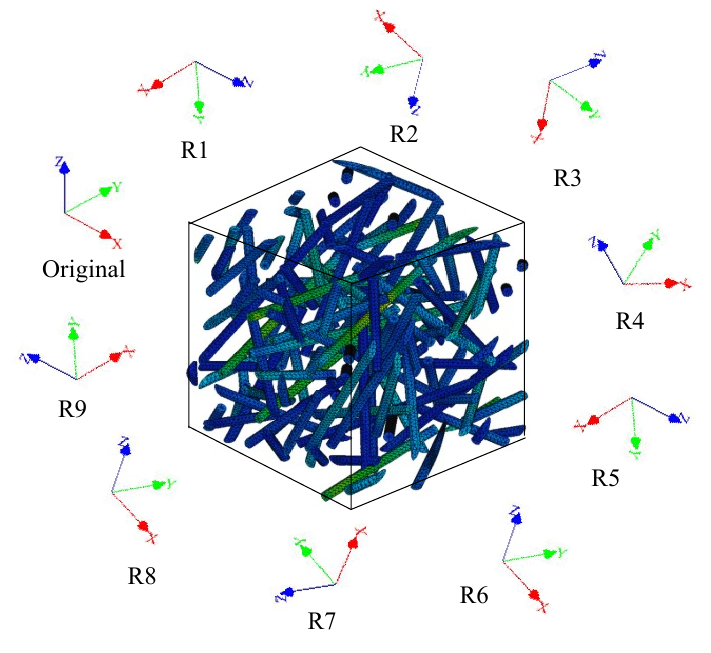}
\caption{An illustration showing the data augmentation approach considering multiple configurations and their corresponding coordinate systems. The RVE contour plot is taken from \cite{Mirkhalaf2022}.}
\label{rotation_coo}
\end{figure}

\begin{figure*}[h!]
% Starting font size 12
% Answer: [trim={left bottom right top},clip]
\centering
\includegraphics[scale=0.8, trim={0 0 1cm 0},clip]
{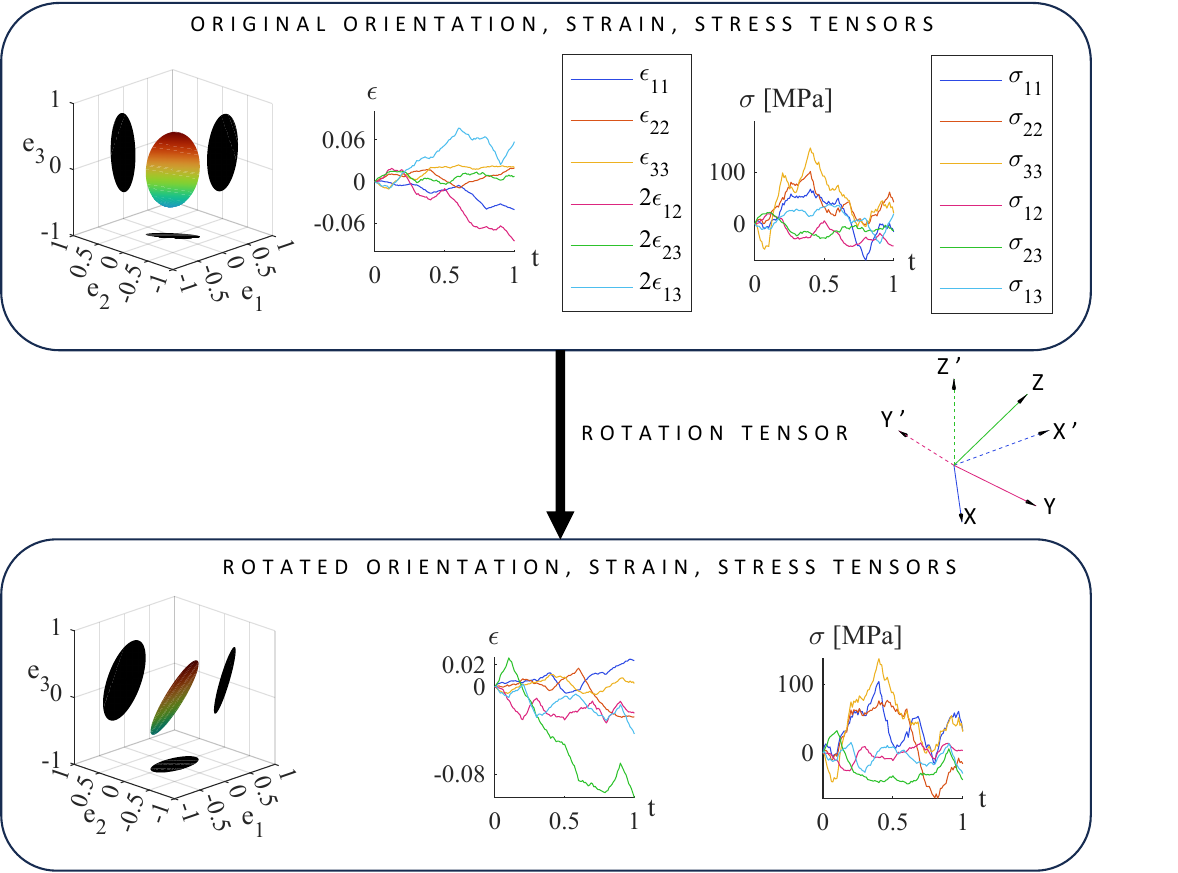}
\caption{A representation of the proposed data augmentation approach using random rotations. The second order orientation, strain, and stress tensors are rotated from one configuration to another configuration using a rotation tensor relating the corresponding coordinate systems.}
\label{rotation_illustration}
\end{figure*}

Instead of increasing the dataset with conducting more expensive simulations, this study proposes to augment data using the original dataset. This approach includes rotation of the 6D input and output data to multiple confiduratons using randomized rotation tensors. This is schematically shown in Figure \ref{rotation_coo}. 
Using this approach, it is possible to investigate whether data augmentation of multi-scale micro-mechanical simulations, is a feasible strategy to increase the dataset size, to enhance the RNN prediction capabilities while capturing the non-linear elasto-plastic response. 
Therefore, the training dataset, including the orientation tensor, strain path and stress evolution, was augmented by using fast random rotations, by implementing the Arvo’s \cite{Arvo1992} algorithm. Each second order tensor, \ie the orientation, the strain, and the stress, can be represented by a 3 $\times$ 3 matrix:
\begin{equation}
    \boldsymbol{a} = \begin{bmatrix} a_{11} & a_{12} & a_{13} \\ & a_{22} & a_{23} \\ 
    \phantom{a_{13}} & \phantom{a_{23}} & a_{33} \\ 
    \end{bmatrix},
    \boldsymbol{\varepsilon} = \begin{bmatrix} \varepsilon_{11} & \varepsilon_{12} & \varepsilon_{13} \\ 
    & \varepsilon_{22} & \varepsilon_{23} \\ 
    \phantom{\varepsilon_{13}} & \phantom{\varepsilon_{23}} & \varepsilon_{33} \\ 
    \end{bmatrix}
    ,
    \boldsymbol{\sigma} = \begin{bmatrix} \sigma_{11} & \sigma_{12} & \sigma_{13} \\ 
    & \sigma_{22} & \sigma_{23} \\ 
    \phantom{\sigma_{13}} & \phantom{\sigma_{23}} & \sigma_{33} \\ 
    \end{bmatrix}.
\end{equation}
In the Arvo’s algorithm, the rotation matrix ($\textbf{\textit{R}}$) is defined as follows:
\begin{equation}
    \boldsymbol{R} = \begin{bmatrix}
    \cos(\theta) & \sin(\theta) & 0 \\
    -\sin(\theta) & \cos(\theta) & 0 \\
    0 & 0 & 1 
    \end{bmatrix}, \quad \text{for } \theta \in [0, 2\pi].
\end{equation}
A unit vector ($\textbf{\textit{v}}$), parallel to the plane of reflection, and the reflection is given by the negatively scaled Householder matrix ($-\textbf{\textit{H}}$):
\begin{equation}
   \boldsymbol{v} = \begin{bmatrix}
    \cos(\phi)\sqrt{z} \\
    \sin(\phi)\sqrt{z} \\
    \sqrt{1-z}
    \end{bmatrix}, \quad \text{for } \phi\in [0, 2\pi], \quad \text{and } z \in [0, 1] ,
\end{equation}
\begin{equation}
    -\boldsymbol{H} = 2\boldsymbol{vv}^T - \boldsymbol{I}.
\end{equation}
A random rotation matrix ($\textbf{\textit{M}}$), is defined by
\begin{equation}
    \boldsymbol{M} = -\boldsymbol{HR}.
\end{equation}
The random rotation matrix ($\textbf{\textit{M}}$) was applied to the strain, orientation, and stress tensors:
\begin{equation}
    \begin{Bmatrix} \boldsymbol{a}_{r}\\ \boldsymbol{\sigma}_{r}\\ \boldsymbol{\varepsilon}_{r}\\ \end{Bmatrix} = \boldsymbol{M} \cdot \begin{Bmatrix} \boldsymbol{a}\\ \boldsymbol{\sigma}\\ \boldsymbol{\varepsilon}\\ \end{Bmatrix} \cdot \boldsymbol{M}^{T}.
\end{equation}
Effectively, the coordinate frame of the training data has been randomly rotated in the 3D space, as illustrated in Figure \ref{rotation_illustration}.

%#########################################################################################
\section{RNN Model Development}
\label{RNN model development}
An RNN model architecture was chosen for training and testing, to evaluate the data augmentation approach. The initial RNN model implemented in this study was first developed by Friemann et al. \cite{Friemann2023} and further developed by Cheung and Mirkhalaf \cite{Cheung2024}.

\subsection{Neural network model architecture}
% The possibility to train the developed RNN model on the original and augmented data was investigated. 
The RNN has 13 inputs, comprising of 6 unique orientation tensor components, a sequence of 6 strain tensor components, and a fiber volume fraction. The output of the RNN is a sequence of 6 unique stress tensor components. Thus, ensuring that complex 6-dimensional stress-strain evolutions could be generated. 

\begin{figure*}[h!]
% Answer: [trim={left bottom right top},clip]
% Starting font size 12
\centering
\includegraphics[scale=0.5]{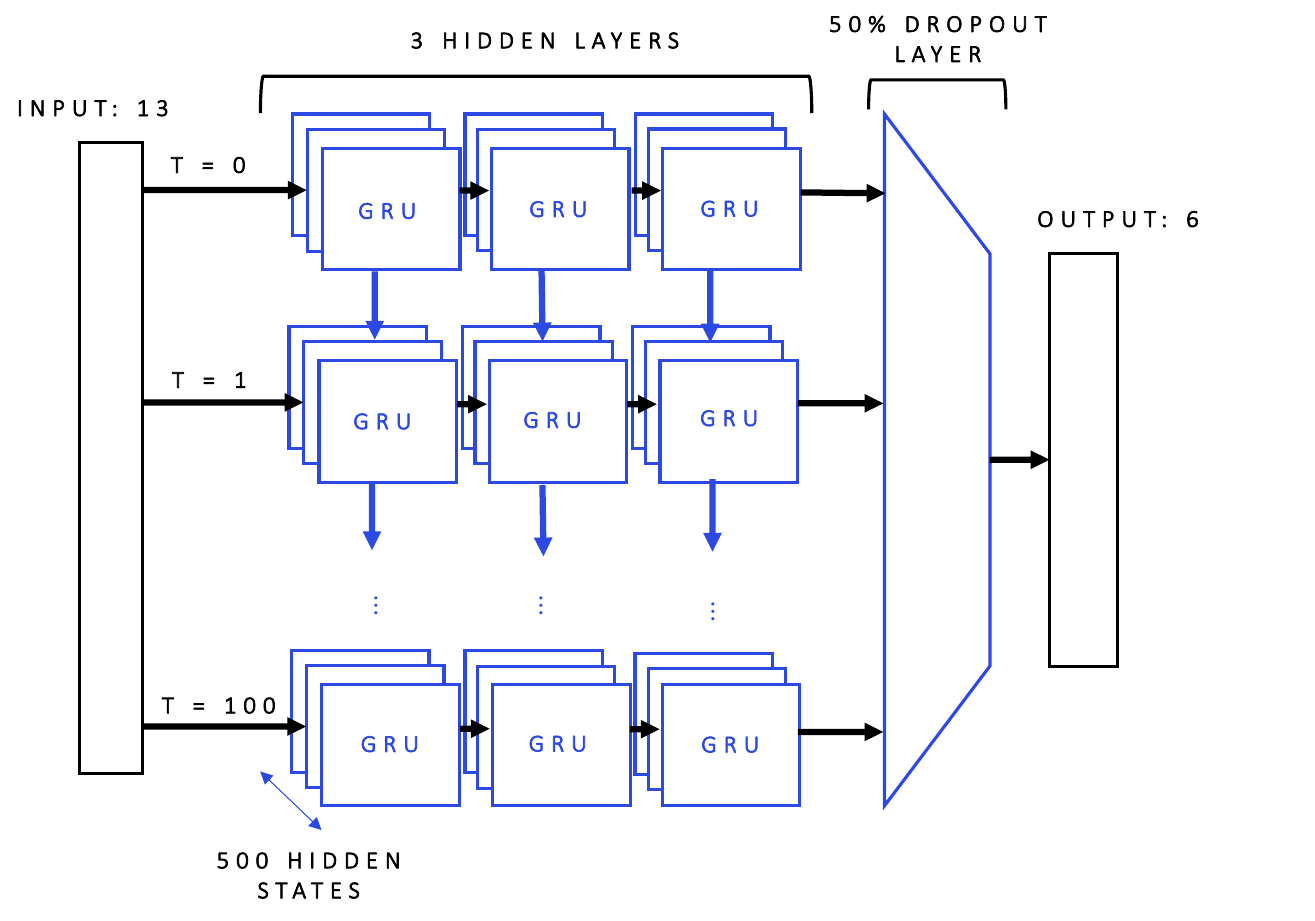}
\caption{The RNN architecture in which, each GRU unit contains 500 hidden states.}
\label{RNN architecture}
\end{figure*}

The RNN architecture was composed of three Gated Recurrent Unit (GRU) layers \cite{Cho2014}, each with 500 hidden states. The GRU updates for the next time input. In this study, 100 time steps were utilized, however, the RNN architecture is not limited to any number of time steps. Following the GRU layers, there exist a dropout layer \cite{Srivastava2014} with a 50\% dropout rate. After that, there is the final layer including 6 neurons (for the 6 output stress components).
 Figure \ref{RNN architecture} illustrates the RNN architecture.

\subsection{Training of the neural network}
\noindent
The original data was divided into a training, validation, and test datasets including 80\%, 15\% and 5\% of the data, respectively. The training dataset was limited due to the computational effort required to produce full-field FE/FFT-simulations. The RNN model was initially trained on the original dataset of 547 data samples, each including 100 time steps. Each sample was randomly rotated, in the range from 1-20 times, effectively increasing the dataset by the number of rotations added.

To effectively train the neural network, the default Matlab loss function for time-series regression was utilized in this study. This function incorporates the sequence length ($S$), the number of outputs ($R$), the target ($t$), and the network prediction ($O$): 
\begin{equation}
    \text{{loss}} = \frac{1}{2S} \sum_{i=1}^{S} \sum_{j=1}^{R} (t_{ij} - O_{ij})^2.
\end{equation}
The loss function was minimized using ADAM optimizer. Default values of parameters such as gradient decay factor $(\beta_{1}=0.9)$, squared gradient decay factor $(\beta_{2}=0.999)$, and the offset $(\varepsilon=1e-8)$ were chosen in a suitable range of neural networks \cite{Kingma2017}. L2-regularization and gradient clipping were also incorporated to prevent overfitting and exploding gradients respectively \cite{Ying2019, Pascanu2013}.

Optimizing hyperparameters is critical to effectively train the neural networks. Particularly, this includes maximum epochs, minimum batch size, initial learning rate, learning rate drop period and factor, and gradient threshold. These are optimized based on learning rate decay in relation to the number of iterations. Bayesian optimization function \cite{Snoek2012} in Matlab was incorporated to optimize the hyperparameters, allowing for up to 65 trials with the objective of minimizing validation loss. The final optimized parameters were determined by selecting the iteration with the lowest validation loss from the best trial.

In this study, training hyperparameters were optimized initially using the original dataset. They were iteratively checked as more rotations were added to the dataset yet did not add significant improvements to the performance. Therefore, in the range from zero rotations to 15 rotations, the hyperparameters were kept constant, \ie identical to the original data. Finally, they were additionally optimized for 15 and 20 rotations.

%#########################################################################################
\section{Results \& Discussion}
\label{Results}
The trained RNN was evaluated based on its predictive capabilities in capturing the non-linear elasto-plastic responses of SFRCs, with various amount of augmented data. Subsequently, the implementation of data augmentation via random rotations was evaluated. Following this, specific loading cases, such as uniaxial, biaxial stress, and plane strain loadings, were examined to assess the RNN's ability to predict non-random stress-strain paths.

\subsection{Evaluation metrics}
The trained RNN was evaluated based on how accurate it could predict the von Mises stress, as given in Equation (\ref{von_mises}): 
\begin{equation}
\label{von_mises}
\resizebox{0.9\hsize}{!}{$
\sigma_V = \sqrt{\frac{1}{2}\left[(\sigma_{11} - \sigma_{22})^2 + (\sigma_{22} - \sigma_{33})^2 + (\sigma_{33} - \sigma_{11})^2\right] + 3(\sigma_{12}^2 + \sigma_{13}^2 + \sigma_{23}^2)}
$}.
\end{equation}
From this, the mean relative error (MeRE) and the maximum relative error (MaRE) were calculated: 
\begin{equation}
\label{MeRE}
\text{MeRE} = \frac{{\sqrt{\sum_{t=1}^{T} ( \sigma_t^V - \hat{\sigma}_t^V)^2}}}{{\max(\sigma_t^V)T}},
\end{equation}

\begin{equation}
\label{MaRE}
\text{MaRE} = \frac{{\max(\sigma_t^V - \hat{\sigma}_t^V)}}{{\max(\sigma_t^V)}}.
\end{equation}
The von Mises stress was chosen as a metrics because it allows for an evaluation that incorporates all 6 stress components. Furthermore, the von Mises stress is physically relevant for determining the material’s behavior, providing a meaningful metrics for evaluating the accuracy of the network predictions.

\subsection{Testing results for randomly evolving strain paths}
\begin{figure}[h!]
\centering
\includegraphics[width=0.70\linewidth, trim={0 0 0 11},clip]{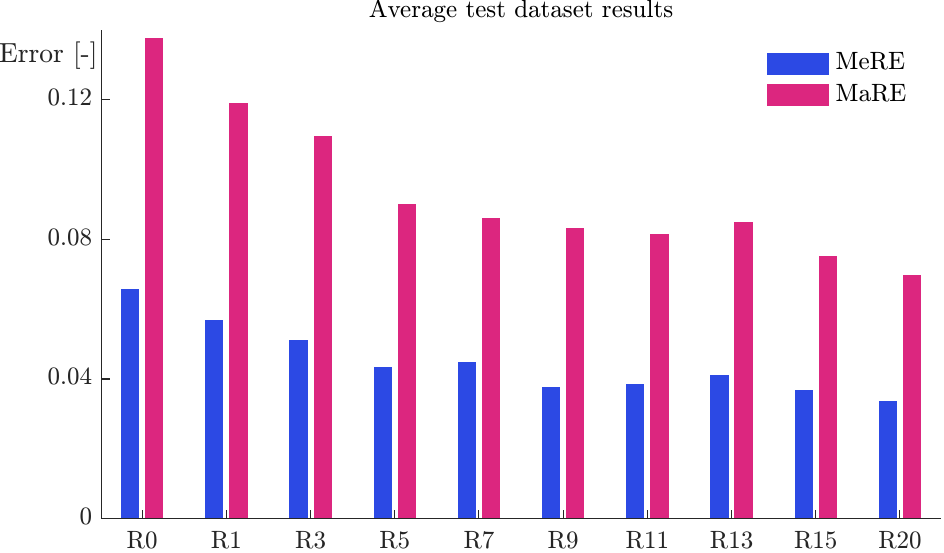}
\caption{Test dataset results for networks trained with different datasets (original and augmented ones). The bars represent the average MeRE and MaRE of the von Mises stress by comparing the networks predictions with the FE/FFT simulations.}
\label{test_dataset}
\end{figure}

The RNN model was initially evaluated for its accuracy in predicting the stress evolutions of the test dataset, \ie the dataset including 5\% of the original data. A consistent decrease in both MeRE and MaRE was observed as the number of augmented datasets increased from the original data to 20 times augmented datasets,  as illustrated in Figure \ref{test_dataset}. 

\begin{figure}[h!]
  \centering
    \includegraphics[scale=0.8]{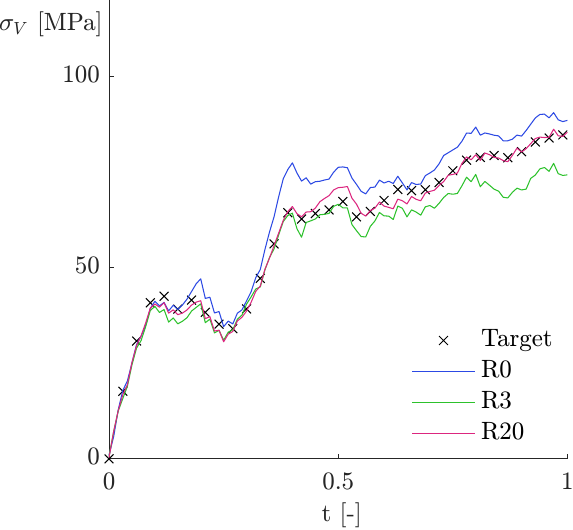}
  \caption{Results of one of the test dataset samples for networks trained with original, 3 augmented, and 20 augmented datasets, showing the von Mises stress calculated from network prediction compared with the FE/FFT simulations.}
  \label{test_ex}
\end{figure}

Data augmentation, achieved by adding 20 randomly rotated stress-strain paths, reduced the MeRE by almost 50\%, from 0.0659 to 0.03398. Therefore, the RNN effectively predicted complex anisotropic non-linear elasto-plastic deformations. An example figure of the predicted random path in relation to the high-fidelity simulated data, is given in Figure \ref{test_ex}.
It can be clearly seen that the augmented datasets are improving the network predictions. 

% The results of this study highlights that data augmentation is a novel approach that significantly enhances the prediction capabilities of the RNN surrogate model, achieving a high degree of accuracy.

\subsection{Specific loading tests results}
\begin{figure}[h!]
\centering
\includegraphics[width=0.70\linewidth, trim={0 0 0 11},clip]{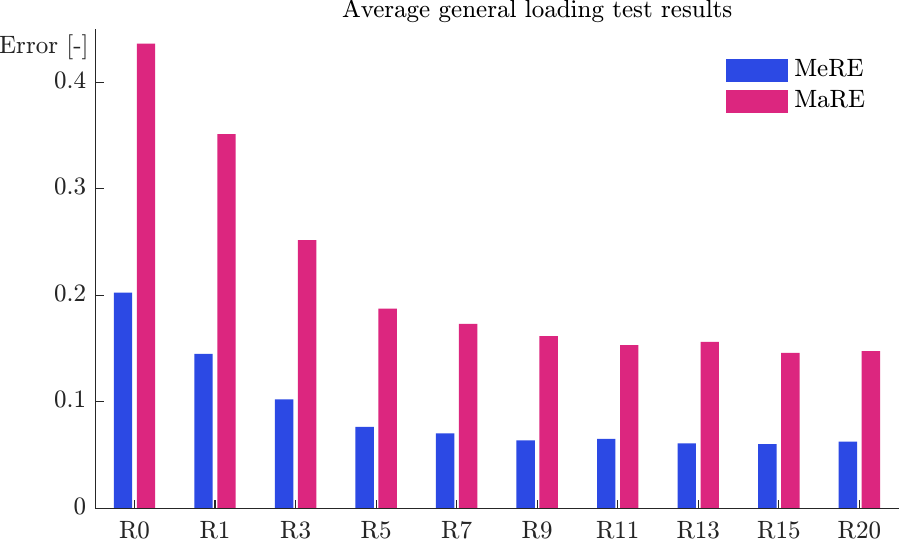}
\caption{Specific loading test results for networks trained with different datasets (original and augmented ones). The bars represent the MeRE and MaRE of the von Mises stress by comparing the networks predictions with the micro-mechanical simulations.}
\label{general_loading}
\end{figure}

The trained RNNs were also tested on specific loading cases, particularly cyclic loading cases of uniaxial stress, biaxial stress and plane strain. This type of loading is commonly applied to materials to assess their performance. However, it is fundamentally different compared to random strain paths, as some stress or strain components are consistently set to zero. The MeRE and MaRE for specific loading tests are shown in Figure \ref{general_loading}.
Once again, the RNN model exhibited accurate predictive capabilities while using augmented datasets. The MeRE was drastically decreased by the data augmentation approach, from 0.20272 to 0.0628. Thus, expanding the dataset resulted in a substantial reduction in prediction errors, aligning more closely with high-fidelity FE/FFT-simulated data.

\begin{figure}[h!]
\centering
\includegraphics[scale=0.8]{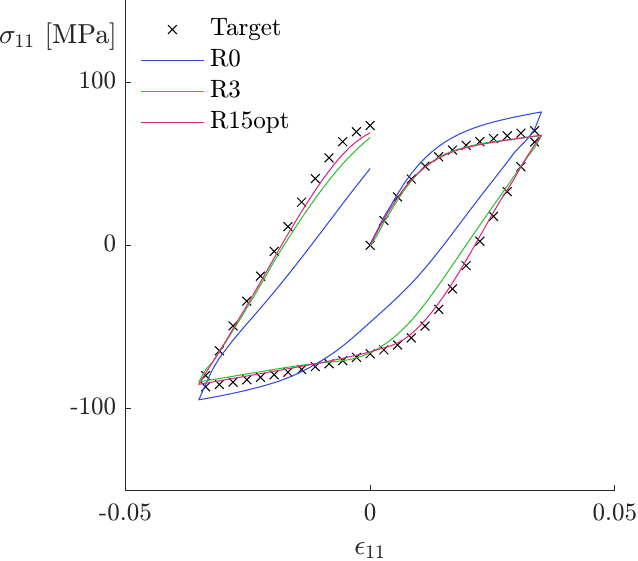}
% Starting font size 12
\caption{Stress-strain plot illustrating the results of a uniaxial stress  $(\sigma_{11})$ loading test on sample 3, for networks trained with original, 3 augmented, and 20 augmented datasets compared with FE/FFT simulations.}
\label{uni_fig}
\end{figure}

\begin{figure}[h!]
\centering
% Starting font size 1
\includegraphics[scale=0.8]{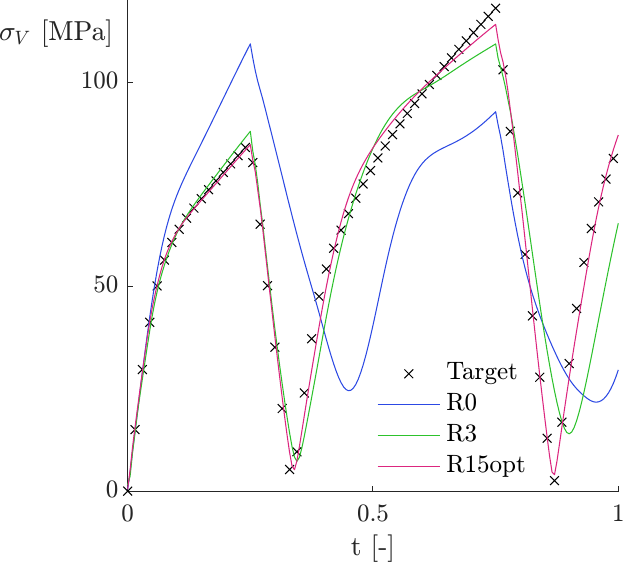}
\caption{Results of a specific loading test on sample 5 under biaxial stress loading $(\sigma_{11}+\sigma_{22})$ for networks trained with the original, 3 augmented, and 20 augmented datasets, showing the von Mises stress calculated from networks predictions compared to micro-mechanical simulations.}
\label{biaxial_fig}
\end{figure}

Furthermore, the RNNs stress-strain predictions together with the original micro-mechanical simulations for a uniaxial stress cyclic load is given in Figure \ref{uni_fig}. 
The network trained on the original dataset poorly matches the shape of the cyclic loading curve. However, as more rotations are added, the predicted output progressively aligns closely with the simulated data. Similarly, for a biaxial normal stress loading case ($\sigma_{11}$ and $\sigma_{22}$), a load cycle is analyzed. From the output stress components, the von Mises stress is calculated and given in Figure \ref{biaxial_fig}.

% This demonstrates that additional rotations enhance the predictive capabilities of the RNN. However, the neural network is poor at predicting the dimensions in which zero stress was applied. Thus, this is a novel approach to augment simulated high-fidelity data in 6D space.

% However, it remains an open question, whether random rotations are optimal for this approach. Since the strain paths were randomly generated in the 6D space, there could be a set of rotations, which better cover the 6D space. Nonetheless, the novel approach of augmenting data through rotations proved to be a feasible strategy for increasing the dataset size and improving the RNN's accuracy. 

The obtained results in this study show the effectiveness of the proposed data augmentation method. It can dramatically reduce the required computational/experimental cost for developing data-driven models. For instance, producing 3D models using experimental data would be challenging since typical testing rigs only allow for specific loading scenarios \cite{LIU2021}. However, fatigue or other mechanical behaviors of SFRCs are crucial for accurate modeling, yet are typically unsupported by classical models \cite{Jain2016}. Experimental results may be utilized to discover unknown constitutive laws. For example, surrogate models trained on experimental data have demonstrated improved predictions of strain hardening in titanium under uniaxial stress \cite{Tasdemir2023}. By implementing test-rig setups, one may produce stress-strain paths in a limited number of directions. Such a limited dataset could then be easily expanded using the proposed data augmentation approach. By allowing a network to train on experimental data, it allows for modeling of unknown laws and facilitate better predictions of the material behavior. This could be crucial for advancing the design of composites and other materials. Thus, this approach to augment data has valuable potentials for advancements in a wide range of industries. It should also be mentioned that despite the obtained great results, the method could potentially be improved by choosing the rotations angles in a \textit{systematic} way. In other words, whether or not \textit{random} rotations are optimal for this approach remains as an open question and requires further investigations. 

%#########################################################################################
%\section{Discussion}
%\label{Discussion}

%#########################################################################################
\section{Conclusions}
\label{Conclusion}
\noindent
Macroscopic behavior of different composite materials, including SFRCs, depends on micro-structural parameters. Therefore, establishing a structure-property relationship requires the use of micro-mechanical models. By using a computational homogenization method applied to realistic RVEs, mimicking the actual material micro-structure, highly accurate predictions are obtained. Nonetheless, difficult RVE generations and computationally expensive simulations remain as main challenges. More recently, data-driven methods, using ANNs, have been proposed as an alternative method to solve these issues. Yet, the data-hungry nature of ANNs poses a challenge for generation of required data for training and validation of an ANN model, which limits advancements of data-driven models. 

%Traditionally, computational simulations have been used to cover 3D models of anisotropic strain hardening. Previously, RNNs have been implemented to predict path-dependent plasticity, on mean-field models and fined-tuned using full-field models. Yet, despite advances, the data-hungry nature of ANNs, especially RNNs, remains a challenge. To the best knowledge of the authors in this paper, no previous 3D micro-mechanical models, have exclusively been trained on full field simulations. 

In this study, we addressed the challenge of limited high-fidelity data for training ANNs by introducing a novel approach - augmenting the original dataset through rotations. % The study expanded the path-dependent time series of anisotropic strain hardening data by rotating the original data, significantly increasing data size without additional simulations.
Using the proposed method, the original dataset of limited high-fidelity simulations was expanded to varying extents by using different amounts of random rotations. An RNN was trained and validated using different datasets (the original dataset and different augmented datasets) of path-dependent non-linear elasto-plastic behavior of SFRCs. The results demonstrated that the proposed data augmentation approach significantly mitigated the data requirement for deep-learning-enhanced modeling of SFRCs.

% An RNN was developed as a surrogate model for the non-linear elasto-plastic response of SFRC microstructures. 

% The results demonstrated accurate prediction capabilities of the RNN, as it could replicate complex loading-paths. Consequently, this study shows an original approach to produce unoriginal data, for datasets of any dimensional size.
We believe that the proposed data augmentation approach is not exclusive to SFRCs and may be used for non only other composites,  but also other kind of materials such as polymers, metals, ceramics etc. Also, the data augmentation method could potentially be applied to lower scale models such as Molecular Dynamics simulations to develop efficient surrogate models. This can dramatically reduce the required time and computational resources for dataset developments, and hence, results in accurate and remarkably efficient data-driven models for modeling and designing materials. 

% In conclusion, this study addresses a critical challenge in modeling of SFRCs, by introducing a novel data augmentation technique using rotations. The developed RNN demonstrates promising accuracy in predicting complex material responses, enabling advancements in computational modeling of SFRCs. It also contributes to the field of material science, computational mechanics, and data driven modeling. Thus, this study highlights the importance of innovative strategies in overcoming data limitations for effective ANN-based modeling.

%#########################################################################################
\section*{Acknowledgment}
S.M. Mirkhalaf and P. Uvdal gratefully acknowledge financial support from the Swedish Research Council (VR grant: 2019-04715) and the University of Gothenburg.

%#########################################################################################
\bibliographystyle{unsrt}
%\bibliography{2.References}{}

\begin{thebibliography}{10}

\bibitem{Schneider2016}
M.~Schneider, F.~Ospald, and M.~Kabel.
\newblock Computational homogenization of elasticity on a staggered grid.
\newblock {\em International Journal for Numerical Methods in Engineering},
  105(9):693--720, 2016.
\newblock cited By 86.

\bibitem{Spahn2014}
J.~Spahn, H.~Andrä, M.~Kabel, and R.~Müller.
\newblock A multiscale approach for modeling progressive damage of composite
  materials using fast fourier transforms.
\newblock {\em Computer Methods in Applied Mechanics and Engineering},
  268:871--883, 2014.

\bibitem{Hagnell2019}
M.~K. Hagnell and M.~Åkermo.
\newblock The economic and mechanical potential of closed loop material usage
  and recycling of fibre-reinforced composite materials.
\newblock {\em Journal of Cleaner Production}, 223:957--968, 6 2019.

\bibitem{Jain2016}
Atul Jain, Jose~M. Veas, Stefan Straesser, Wim~Van Paepegem, Ignaas Verpoest,
  and Stepan~V. Lomov.
\newblock The master sn curve approach – a hybrid multi-scale fatigue
  simulation of short fiber reinforced composites.
\newblock {\em Composites Part A: Applied Science and Manufacturing},
  91:510--518, 12 2016.

\bibitem{Mortazavian2015}
Seyyedvahid Mortazavian and Ali Fatemi.
\newblock Fatigue behavior and modeling of short fiber reinforced polymer
  composites: A literature review.
\newblock {\em International Journal of Fatigue}, 70:297--321, 1 2015.

\bibitem{Tikarrouchine2018}
E.~Tikarrouchine, G.~Chatzigeorgiou, F.~Praud, B.~Piotrowski, Y.~Chemisky, and
  F.~Meraghni.
\newblock Three-dimensional fe2 method for the simulation of non-linear,
  rate-dependent response of composite structures.
\newblock {\em Composite Structures}, 193:165--179, 6 2018.

\bibitem{Selmi2011}
A.~Selmi, I.~Doghri, and L.~Adam.
\newblock Micromechanical simulations of biaxial yield, hardening and plastic
  flow in short glass fiber reinforced polyamide.
\newblock {\em International Journal of Mechanical Sciences}, 53(9):696--706,
  2011.
\newblock cited By 16.

\bibitem{Tian2015}
W.~Tian, L.~Qi, J.~Zhou, J.~Liang, and Y.~Ma.
\newblock {Representative volume element for composites reinforced by spatially
  randomly distributed discontinuous fibers and its applications}.
\newblock {\em Composite Structures}, 131:366--373, 2015.

\bibitem{Mirkhalaf2020}
S.M. Mirkhalaf, E.H. Eggels, T.J.H. {van Beurden}, F.~Larsson, and
  M.~Fagerström.
\newblock A finite element based orientation averaging method for predicting
  elastic properties of short fiber reinforced composites.
\newblock {\em Composites Part B: Engineering}, 202:108388, 2020.

\bibitem{Qi2015}
L.~Qi, W.~Tian, and J.~Zhou.
\newblock {Numerical evaluation of effective elastic properties of composites
  reinforced by spatially randomly distributed short fibers with certain aspect
  ratio}.
\newblock {\em Composite Structures}, 131:843--851, 2015.

\bibitem{Hoang2016}
T.~H. Hoang, M.~Guerich, and J.~Yvonnet.
\newblock Determining the size of rve for nonlinear random composites in an
  incremental computational homogenization framework.
\newblock {\em Journal of Engineering Mechanics}, 142, 5 2016.

\bibitem{Harper2012}
L.~T. Harper, C.~Qian, T.~A. Turner, S.~Li, and N.~A. Warrior.
\newblock Representative volume elements for discontinuous carbon fibre
  composites - part 2: Determining the critical size.
\newblock {\em Composites Science and Technology}, 72:204--210, 1 2012.

\bibitem{Bargmann2018}
S.~Bargmann, B.~Klusemann, J.~Markmann, J.E. Schnabel, K.~Schneider,
  C.~Soyarslan, and J.~Wilmers.
\newblock Generation of 3d representative volume elements for heterogeneous
  materials: A review.
\newblock {\em Progress in Materials Science}, 96:322 -- 384, 2018.

\bibitem{Pan2008}
Yi~Pan, Lucian Iorga, and Assimina~A. Pelegri.
\newblock Numerical generation of a random chopped fiber composite rve and its
  elastic properties.
\newblock {\em Composites Science and Technology}, 68:2792--2798, 10 2008.

\bibitem{Mirkhalaf2019-ICCM}
S.M. Mirkhalaf, E.H. Eggels, A.T. Anantharanga, F.~Larsson, and M.~Fagerström.
\newblock Short fiber composites: Computational homogenization vs orientation
  averaging.
\newblock {\em ICCM22 2019}, page 3000, 2019.

\bibitem{Mozaffar2019}
M.~Mozaffar, R.~Bostanabad, W.~Chen, K.~Ehmann, J.~Cao, and M.~A. Bessa.
\newblock Deep learning predicts path-dependent plasticity.
\newblock {\em Proceedings of the National Academy of Sciences},
  116(52):26414--26420, December 2019.

\bibitem{Mentges2021}
N.~Mentges, B.~Dashtbozorg, and S.~M. Mirkhalaf.
\newblock A micromechanics-based artificial neural networks model for elastic
  properties of short fiber composites.
\newblock {\em Composites Part B: Engineering}, 213, 5 2021.

\bibitem{LIU2021}
Xin Liu, Su~Tian, Fei Tao, and Wenbin Yu.
\newblock A review of artificial neural networks in the constitutive modeling
  of composite materials.
\newblock {\em Composites Part B: Engineering}, 224:109152, 2021.

\bibitem{Friemann2023}
J~Friemann, B~Dashtbozorg, M~Fagerstr{\"o}m, and SM~Mirkhalaf.
\newblock A micromechanics-based recurrent neural networks model for
  path-dependent cyclic deformation of short fiber composites.
\newblock {\em International Journal for Numerical Methods in Engineering},
  124(10):2292--2314, 2023.

\bibitem{Ghane2023}
E~Ghane, M~Fagerstr{\"o}m, and SM~Mirkhalaf.
\newblock A multiscale deep learning model for elastic properties of woven
  composites.
\newblock {\em International Journal of Solids and Structures}, 282:112452,
  2023.

\bibitem{Bonatti2022}
Colin Bonatti and Dirk Mohr.
\newblock On the importance of self-consistency in recurrent neural network
  models representing elasto-plastic solids.
\newblock {\em Journal of the Mechanics and Physics of Solids}, 158:104697,
  2022.

\bibitem{Liu2023}
Xiao Liu, Ji~He, and Shiyao Huang.
\newblock Mechanistically informed artificial neural network model for
  discovering anisotropic path-dependent plasticity of metals.
\newblock {\em Materials and Design}, 226, 2 2023.

\bibitem{Cheung2024}
Hon~Lam Cheung and Mohsen Mirkhalaf.
\newblock A multi-fidelity data-driven model for highly accurate and
  computationally efficient modeling of short fiber composites.
\newblock {\em Composites Science and Technology}, In Press, 2024.

\bibitem{Lecun2015}
Yann Lecun, Yoshua Bengio, and Geoffrey Hinton.
\newblock Deep learning.
\newblock {\em Nature}, 521:436--444, 5 2015.

\bibitem{Agrawal2019}
Ankit Agrawal and Alok Choudhary.
\newblock Deep materials informatics: Applications of deep learning in
  materials science.
\newblock {\em MRS Communications}, 9:779--792, 9 2019.

\bibitem{Heidenreich2023}
Julian~N. Heidenreich, Colin Bonatti, and Dirk Mohr.
\newblock Transfer learning of recurrent neural network‐based plasticity
  models.
\newblock {\em International Journal for Numerical Methods in Engineering}, 9
  2023.

\bibitem{Ghane2023-RNN}
Ehsan Ghane, Martin Fagerström, and Mohsen Mirkhalaf.
\newblock Recurrent neural networks and transfer learning for elasto-plasticity
  in woven composites.
\newblock {\em Pre-print at: https://doi.org/10.48550/arXiv.2311.13434}, 2023.

\bibitem{JUNG2022}
Jiyoung Jung, Yongtae Kim, Jinkyoo Park, and Seunghwa Ryu.
\newblock Transfer learning for enhancing the homogenization-theory-based
  prediction of elasto-plastic response of particle/short fiber-reinforced
  composites.
\newblock {\em Composite Structures}, 285:115210, 2022.

\bibitem{Advani1987}
Suresh~G. Advani and Charles~L. Tucker.
\newblock The use of tensors to describe and predict fiber orientation in short
  fiber composites.
\newblock {\em Journal of Rheology}, 31:751--784, 11 1987.

\bibitem{Arvo1992}
James Arvo.
\newblock Iii.4 - fast random rotation matrices.
\newblock In DAVID KIRK, editor, {\em Graphics Gems III (IBM Version)}, pages
  117--120. Morgan Kaufmann, San Francisco, 1992.

\bibitem{Mirkhalaf2016}
S.M. Mirkhalaf, F.M. {Andrade Pires}, and Ricardo Simoes.
\newblock Determination of the size of the representative volume element (rve)
  for the simulation of heterogeneous polymers at finite strains.
\newblock {\em Finite Elements in Analysis and Design}, 119:30--44, 2016.

\bibitem{Mentges2023}
N.~Mentges, H.~Çelik, C.~Hopmann, M.~Fagerström, and S.M. Mirkhalaf.
\newblock Micromechanical modelling of short fibre composites considering fibre
  length distributions.
\newblock {\em Composites Part B: Engineering}, 264:110868, 2023.

\bibitem{Mirkhalaf2022}
S.M. Mirkhalaf, T.J.H. {van Beurden}, M.~Ekh, F.~Larsson, and M.~Fagerström.
\newblock An fe-based orientation averaging model for elasto-plastic behavior
  of short fiber composites.
\newblock {\em International Journal of Mechanical Sciences}, 219:107097, 2022.

\bibitem{Cho2014}
Kyunghyun Cho, Bart van Merri{\"e}nboer, Caglar Gulcehre, Dzmitry Bahdanau,
  Fethi Bougares, Holger Schwenk, and Yoshua Bengio.
\newblock Learning phrase representations using {RNN} encoder{--}decoder for
  statistical machine translation.
\newblock In {\em Proceedings of the 2014 Conference on Empirical Methods in
  Natural Language Processing ({EMNLP})}, pages 1724--1734, Doha, Qatar,
  October 2014. Association for Computational Linguistics.

\bibitem{Srivastava2014}
Nitish Srivastava, Geoffrey Hinton, Alex Krizhevsky, Ilya Sutskever, and Ruslan
  Salakhutdinov.
\newblock Dropout: A simple way to prevent neural networks from overfitting.
\newblock {\em J. Mach. Learn. Res.}, 15(1):1929–1958, jan 2014.

\bibitem{Kingma2017}
Diederik~P. Kingma and Jimmy Ba.
\newblock Adam: {A} method for stochastic optimization.
\newblock In Yoshua Bengio and Yann LeCun, editors, {\em 3rd International
  Conference on Learning Representations, {ICLR} 2015, San Diego, CA, USA, May
  7-9, 2015, Conference Track Proceedings}, 2015.

\bibitem{Ying2019}
Xue Ying.
\newblock An overview of overfitting and its solutions.
\newblock {\em Journal of Physics: Conference Series}, 1168(2):022022, feb
  2019.

\bibitem{Pascanu2013}
Razvan Pascanu, Tomas Mikolov, and Yoshua Bengio.
\newblock On the difficulty of training recurrent neural networks.
\newblock In {\em Proceedings of the 30th International Conference on
  International Conference on Machine Learning - Volume 28}, ICML'13, page
  III–1310–III–1318, 2013.

\bibitem{Snoek2012}
Jasper Snoek, Hugo Larochelle, and Ryan~P Adams.
\newblock Practical bayesian optimization of machine learning algorithms.
\newblock In F.~Pereira, C.J. Burges, L.~Bottou, and K.Q. Weinberger, editors,
  {\em Advances in Neural Information Processing Systems}, volume~25, 2012.

\bibitem{Tasdemir2023}
Burcu Tasdemir, Vito Tagarielli, and Antonio Pellegrino.
\newblock A data-driven model of the yield and strain hardening response of
  commercially pure titanium in uniaxial stress.
\newblock {\em Materials and Design}, 229, 5 2023.

\end{thebibliography}

\newpage
\appendix

\newpage
\listoffigures

\listoftables

\end{document}